# Supremacy by Accelerated Warfare through the Comprehension Barrier and Beyond: Reaching the Zero Domain and Cyberspace Singularity

Jan Kallberg[1]

[1]*United States Military Academy, the Department of Social Sciences, and the Army Cyber Institute at West Point, Spellman Hall, 2101 New South Post Rd, West Point NY 10996, USA*

Corresponding author: Jan Kallberg, United States Military Academy and Army Cyber Institute, Spellman Hall room 4-33, 2101 New South Post Rd, West Point NY 10996, USA, Email: jan.kallberg@usma.edu.

# Supremacy by Accelerated Warfare through the Comprehension Barrier and Beyond: Reaching the Zero Domain and Cyberspace Singularity

"IN THE LAND OF THE BLIND, THE ONE-EYED MAN IS KING."

ERASMUS OF ROTTERDAM, 16TH CENTURY

**Introduction**

It is questionable and even unlikely that cyber supremacy could be reached by overwhelming capabilities manifested by stacking more technical capacity and adding attack vectors. The alternative is to use time as the vehicle to supremacy by accelerating the velocity of the engagements beyond the speed at which the enemy can target, and precisely execute and comprehend the events unfolding. The space created beyond the adversary's comprehension is called the Zero Domain. Military traditionally sees the battle space as land, sea, air, space, and cyber domains. When fighting the battle beyond the adversary's comprehension, no traditional warfighting domain serves as a battle space; it is not a vacuum nor an unclaimed *terra nullius*, but instead the Zero Domain.

In the Zero Domain, cyberspace superiority surfaces as the outfall of the accelerated time and a digital space-separated singularity that benefit the more-rapid actor. The Zero Domain has a time space that is accessible only by the rapid actor and a digital landscape that is not accessible by the slower actor due to the execution velocity in enhanced accelerated warfare. Velocity achieves cyber Anti Access/Area Denial (A2/AD), which can be achieved without active initial interchanges by accelerating the execution and cyber ability in a solitaire state. During this process, any adversarial probing engagements only affect the actor on the approach to the Comprehension Barrier and once arrived in the Zero Domain a complete state of A2/AD is present.

From that point forward, the actor that reached the Zero Domain has cyberspace singularity while the accelerated actor is the only actor who can understand the digital landscape, engage unilaterally without an adversarial ability to counterattack or interfere, and hold the ability to decide when, how, and where to attack. In the Zero Domain, the accelerated singularity forges the battlefield gravity and thrust into a single power that denies adversarial cyber operations and acts as one force of destruction, extraction, corruption, and exploitation of targeted adversarial digital assets.

**Breaking through the Comprehension Barrier**

There is a point along the trajectory of accelerated warfare where only one warfighting nation comprehends what is unfolding and the sees the cyber terrain; it is an upper barrier for comprehension where the acceleration makes the cyber engagement unilateral. The Comprehension Barrier is dependent on its side's abilities, technical maturity, and institutional structure, and the enemies' weaknesses and lack thereof. Adversaries forged in organizational fear cultures and a strict command structure could, even if technically cognizant and competent, be less able to competitively accelerate the warfare than more agile and less technically able opponents. If both fighting parties are engaged in interchanges, the engagements that are accelerating toward the Comprehension Barrier have increased intensity, as they are faster, more forceful, and less restrained when the stress of acceleration degrades the OODA (Observe, Orient, Decide, Act) loop, command, and control. Once one warfighter breaks through the Comprehension Barrier with maintained control, the conflict changes from a contested cyberspace to battle space singularity. The cyber ability in the Zero Domain battle is derived from a single source. At that point, any engagement can affect only the slower party and not the owner of the singularity because the slower attacker is unable to understand the factual battle landscape and target, arrange its resources, and conduct warfare at the velocity that occurs on the other side of the Comprehension Barrier. If we use real-life references, the warfighter beyond the Comprehension Barrier has full access to situation awareness, can see the landscape and target, and can act as if the war occurred under normal conditions, while the slower warfighter that never reached the Comprehension Barrier is floating around weightless and in pure darkness. Accelerated warfare beyond the Comprehension Barrier caused the slower party to be unable to understand, sense, order, and coordinate the operations. When breaking the Comprehension Barrier, the first of the adversary's final points of comprehension is human deliberation, directly followed by pre-authorization and machine learning, and then these final points of comprehension are passed and the more-rapid actor enters the Zero Domain.

**Time and the Lost Space**

In accelerated cyberwar, time is to cyber what combined time and place were for Clausewitz[1] because the Zero Domain has nullified the importance of the other warfighting domains and created parsimonious singularity through the absence of a common battlespace. Space matters only before the Comprehension Barrier is crossed. The traditional concentration of forces—*where* and *when*—is replaced with *then*, because the singularity occurs first in the Zero Domain. As noted strategist Edward N. Luttwak stated, strategies without the ability to execute are pointless exercises.[2] The accelerated warfare beyond the Comprehension Barrier, if reached, will eradicate the influence of the opponent's cyber strategy because singularity in the Zero Domain removes the opponent's ability to execute.

**The Evaporated OODA Loop**

From an operational standpoint, action beyond the Comprehension Barrier will evaporate and nullify the traditional command and control scheme. In general terms, military command and control (C2) follow the steps of observe, orient, decide, and act, as described in the OODA loop developed by John Boyd in the 1960s (Fig. 1).[3] Accelerated warfare beyond the Comprehension Barrier nullifies the adversary's OODA loop because there is nothing that could be accurately observed, no targets to orient toward, a lack of information and situation awareness with which to make a decision, and the ability to act is limited to spurious actions that have no bearing on the unfolding events. The unique tenets of cyber undermine the utility of the OODA loop.[4] The OODA loop requires the ability to assess ongoing events (as in the initial step of "observe"), but under conditions of anonymity, computational speed in cyber execution, and lack of object permanence, the observations feeding the loop are likely to be inaccurate, if not spurious, as acceleration starts. In accelerated warfare, the OODA loop disappears for the slower party in the engagement if the faster actor breaks the Comprehension Barrier. The rapid actor maintains its OODA loop in the Zero Domain, and conversely, if the rapid actor is no longer able to maintain its position in the Zero Domain, the OODA loop will reemerge for the slower actor as the formerly rapid actor is unable to maintain velocity beyond the Comprehension Barrier.

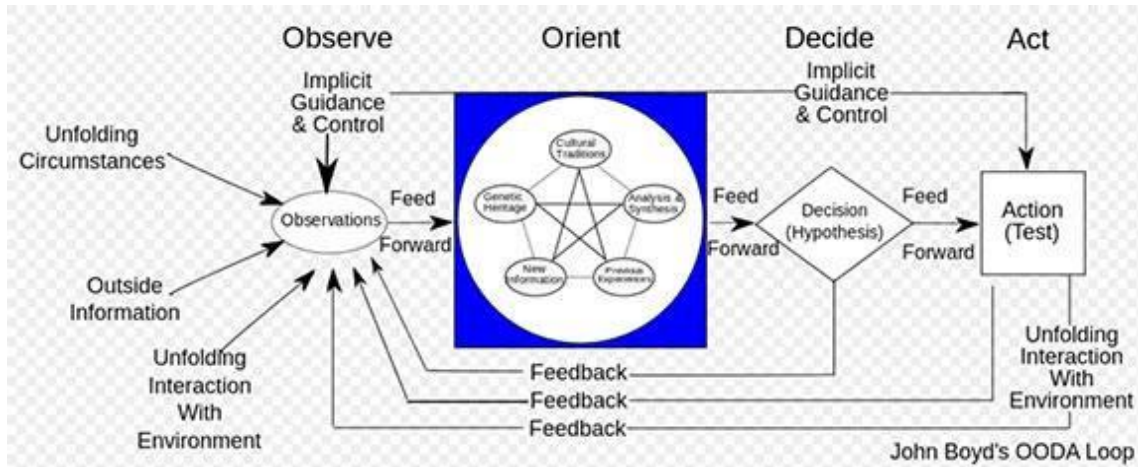

Accelerated warfare beyond the Comprehension Barrier does not seek to be inside the enemy's OODA loop; it removes the OODA loop.

Figure 1. "OODA Boyd" by Patrick Edwin Moran - own work. Licensed under CC BY 3.0 via Wikimedia Commons.

The "orient" stage in the OODA loop—reacting to unfolding events and positioning for a better outcome—assumes a maneuverable space with favorable positions, but the lack of object permanence in cyber brings an ever-changing battlefield and permanent disorientation rather than re-orientation. When these nodes—ever-changing spaces lacking object permanence—are accelerated beyond the Comprehension Barrier, environmental information cannot be structurally understood or ordered outside of the Zero Domain. If the "observe" and "orient" stages are not relevant to the facts of the engagement, then the "decide" stage will fail to deliver the proper course of action and thus lead to an ineffective "act" stage.

Computational speed exacerbates the inability to assess and act, and the increasingly shortened time frames likely to be found in future cyber conflicts will disallow any significant human deliberation. Enemy deliberation, either through leadership or pre-authorization, are ultimately ineffectual once the Comprehension Barrier is passed.

The key to victory historically has been the concept of being able to be inside the opponent's OODA loop, and thereby distort, degrade, and derail the opponent's OODA assessments.[5] In accelerated warfare beyond the Comprehension Barrier, there is no need to be inside the opponent's OODA loop because the accelerated warfare concept removes the OODA loop for the opponent and by doing so disables the opponent's ability to coordinate, seek effect, and command.

**The Zero Domain**

The five traditional battlespace domains are contested spaces (land, sea, air, space, and cyber) where parties interact, engage, have interchanges through which they can structure their understanding of the battle environment to make decisions. Both parties are present in the engagement, and even if one is weaker and less able to challenge, there is a mutual perception of the framing of the fight. The Zero Domain is the battle space beyond the Comprehension Barrier where battle space singularity occurs, and only one actor has access to the OODA loop. The Zero Domain is the warfighting space where accelerated velocity in the warfighting operations removes the enemy's presence. It is the domain with zero opponents. It is not an area denial, because the enemy is unable to accelerate to the level where it could enter the battle space, and it is not access denial because the enemy is not part of the fight once the Comprehension Barrier is broken. Instead, it is a state of cyber A2/AD, but there is no challenge to this state in the Zero Domain because it is an outfall of the establishment of the Zero Domain.

**Short Conclusion**

As a research note, these ideas and concepts are under development and are not the final output. The purpose of the note is to introduce new concepts, open up the discussion, and catalyze comments.

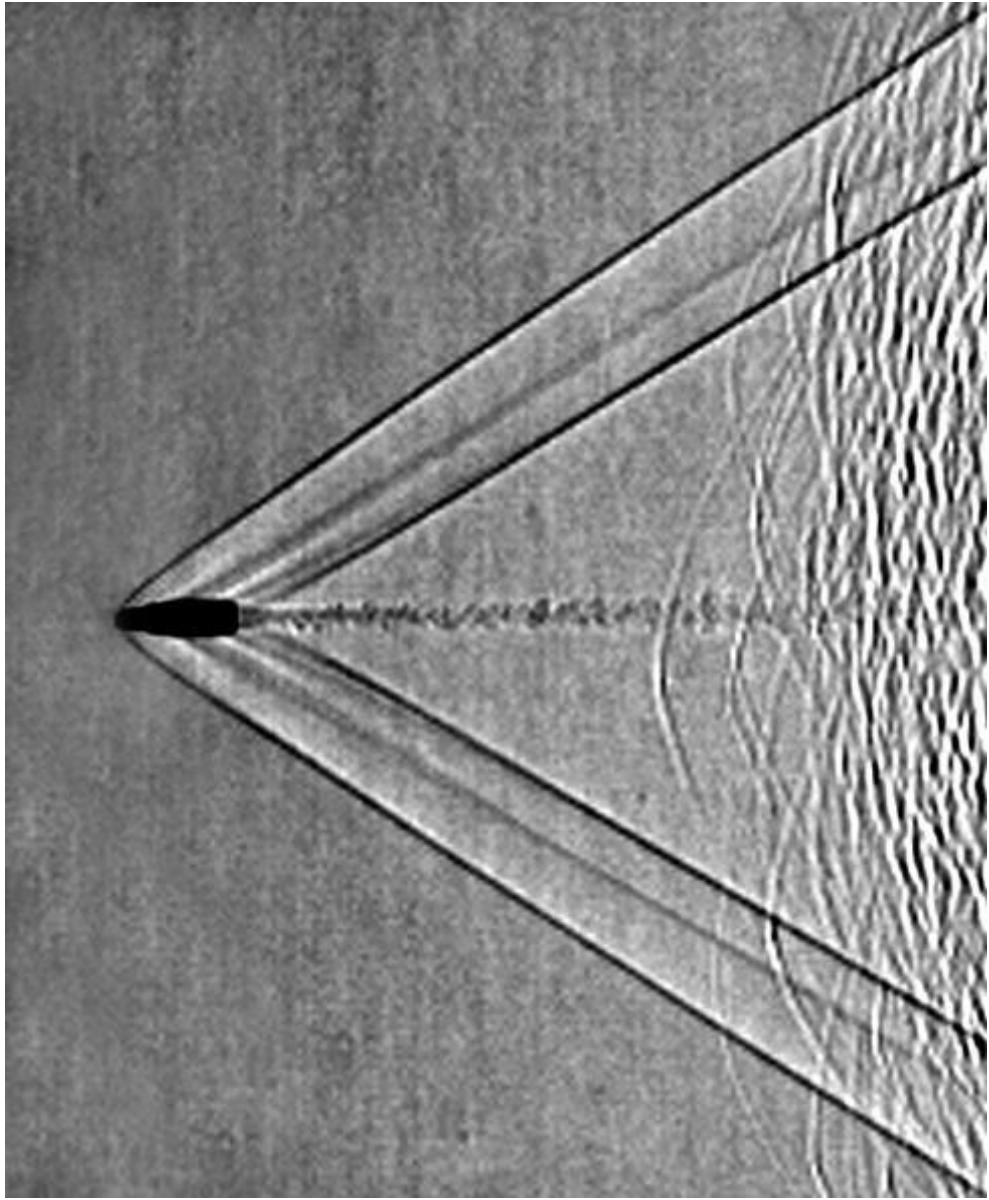

Figure 2. When breaking the Comprehension Barrier, the first of the adversary's final points of comprehension is human deliberation, directly followed by pre-authorization and machine learning, and then these final points of comprehension are passed, and the rapid actor enters the Zero Domain.

By Settles1 [CC BY-SA 3.0 (https://creativecommons.org/licenses/by-sa/3.0)], from Wikimedia Commons.